\documentclass[10pt,twocolumn]{article}

\usepackage{multicol}
\usepackage{graphicx}
\usepackage{amsmath}
\usepackage{lipsum}
\usepackage{color}
\usepackage{siunitx}
\usepackage{authblk}
\usepackage{subfigure}
\usepackage[margin=2cm]{geometry}

\title{Wetting morphologies on randomly oriented fibers}
\author[1]{Alban Sauret}
\author[2]{Fran\c{c}ois Boulogne}
\author[2]{Beatrice Soh}
\author[3]{Emilie Dressaire}
\author[2]{Howard A. Stone}
\affil[1]{Surface du Verre et Interfaces, UMR 125, 93303 Aubervilliers,  France (E-mail: alban.sauret@saint-gobain.com)}
\affil[2]{Department of Mechanical and Aerospace Engineering, Princeton University, Princeton, NJ 08544, USA}
\affil[3]{Department of Mechanical and Aerospace Engineering, New York University Polytechnic School of Engineering, Brooklyn, NY 11201, USA}
\date{ }
\begin{document}

\twocolumn[
    \begin{@twocolumnfalse}
        \maketitle
        \begin{abstract}
            We characterize the different morphologies adopted by a drop of liquid placed on two randomly oriented fibers, which is a first step toward understanding the wetting of fibrous networks. The present work reviews previous modeling for parallel and touching crossed fibers and extends it to an arbitrary orientation of the fibers characterized by the tilting angle and the minimum spacing distance. Depending on the volume of liquid, the spacing distance between fibers and the angle between the fibers, we highlight that the liquid can adopt three different equilibrium morphologies: (1) a column morphology in which the liquid  spreads between the fibers, (2) a mixed morphology where a drop grows at one end of the column or (3) a single drop located at the node. We capture the different morphologies observed using an analytical model that predicts the equilibrium configuration of the liquid based on the geometry of the fibers and the volume of liquid. \\
            \medskip\medskip\medskip
        \end{abstract}
    \end{@twocolumnfalse}
]

\section{Introduction}

The spreading behavior of a liquid placed on a solid substrate controls a broad range of natural and man-made processes, from the clinging of morning dew to spider webs to the coating of surfaces \cite{Contal2004,Brinkmann2004,Chen2007,Zheng2010,Yu2012,Bai2012,White2013}.
Thus, characterizing wetting phenomena offers insights into the complex physics of wet or partially wet systems. These studies also provide knowledge that can be applied to improve and develop industrial methods in which capillary forces play a key role, {\it{e.g.}} coating, mixing and agglomeration \cite{Herminghaus2005,Mitarai2006,Chandra2010,Strauch2012}.

When a volume of liquid is placed between two solid surfaces, a capillary bridge forms. The equilibrium shape of the liquid bridge has been studied for different configurations of the solid surfaces, e.g. flat plates and spherical grains \cite{Rabinovich2005,Restagno2004,DeSouza2008,Wexler2014}. The shape of the liquid bridge minimizes the interfacial free energy. When the distance between the surfaces is increased, the liquid exerts an attractive force that pulls the two surfaces together \cite{Bico2004}. This cohesive capillary force gives rise to the rich mechanical behavior of wet granular matter \cite{Herminghaus2005}.

\begin{figure}
    \centering
\includegraphics[width=8cm]{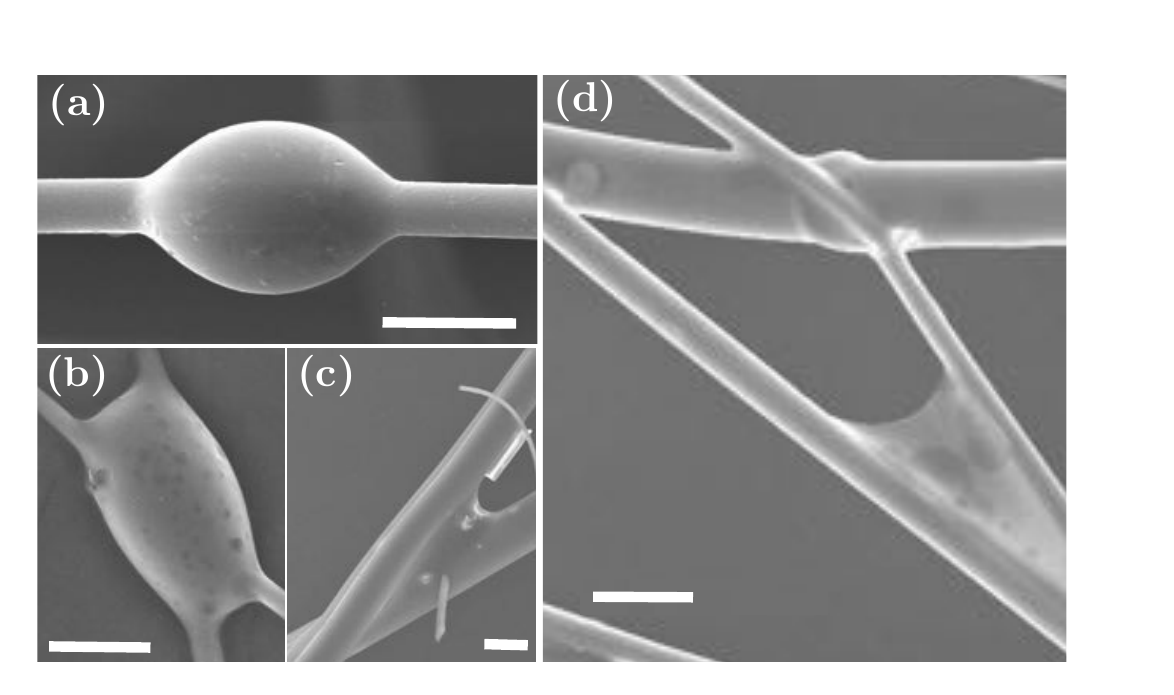}
    \caption{SEM pictures of drops of binder lying on glass wool. (a) Drop on a single fiber, (b) liquid in the drop state between two crossed fibers, (c) and (d) liquid in the column state between two fibers. Scale bars are $10\,\mu{\rm m}$ (Pictures from Saint-Gobain Research, reproduced with permission from Bintein \cite{binteinphd}).}\label{fig:SGR}
\end{figure}

One configuration of solid surfaces has received less attention: the formation of capillary bridges between long cylinders, or fibers \cite{Erle1971,Carroll1976}. It is evident that fibrous media are ubiquitous in both natural systems, such as feathers and hair \cite{Duprat2012}, and engineered products, including paper and textiles \cite{Pezron1995}. Understanding the wetting of fibers is thus important for many industries. The wetting influences the dyeing of textiles, the coloring of human hair and the spreading of ink on paper. In particular, understanding the distribution of liquid in an array of fibers is also critical to the generation of fiber mats used in glass wool for insulation purpose. In this situation the glass fibers are stuck together by a binder fluid. The final properties of the fiber mats are in part controlled by how the wetting binder fluid is distributed among the glass wool before its solidification (figure \ref{fig:SGR}a-d). In addition, glass wool does not swell when in contact with liquid and we will therefore neglect this effect \cite{binteinphd}.

Because of challenges in visualizing the microstructures, fibrous media are complex arrays of fibers that are difficult to study experimentally. Therefore seminal work has focused on the simplest element of an array of fibers: a pair of straight parallel fibers \cite{Princen1970,sauret2015}. These studies have identified two liquid morphologies. A small volume of nonvolatile liquid deposited on a pair of parallel fibers can adopt a hemispherical drop shape or an extended column state. Recently, our work on crossed touching fibers has shown that, in addition to the drop and column states, the liquid can exist in a third morphology: a composite drop/column state referred to as the mixed morphology \cite{Sauret2014}. Analytical models for the shape of the column state on parallel and crossed touching fibers have been previously proposed and compared with experiments.

However, in most fibrous media, the fibers are randomly oriented and spaced, which results in three possible configurations for neighboring pairs of fibers: parallel fibers, touching crossed fibers and non-touching crossed fibers. In considering a global model for fiber arrays, we thus need to account for the latter case, in which the fibers are not touching. Indeed, the closest distance between non-parallel fibers is an additional parameter that affects the equilibrium morphology of the liquid.

In this paper, we study the wetting morphologies on a pair of fibers that are randomly oriented and spaced thus considering a more general situation than previous work performed on liquid bridges between touching or parallel fibers. In particular, we characterize the transitions between the wetting morphologies on a pair of crossed fibers with respect to four variables: the angle between the fibers $\delta$, the distance between the fibers $h$, the fiber radius $a$ and the volume of liquid $V$. Thus, the new model presented in this paper describes the equilibrium wetting morphologies associated with any fiber configurations and recovers the results obtained previously for parallel and touching crossed fibers. We also highlight the understanding of the liquid morphology between randomly oriented fibers in a new 3D diagram. This characterization of equilibrium wetting morphologies is essential for future studies on the properties of wet fibrous media including their drying behaviour.

\begin{figure}
    \centering
   \subfigure[]{\includegraphics[width=6cm]{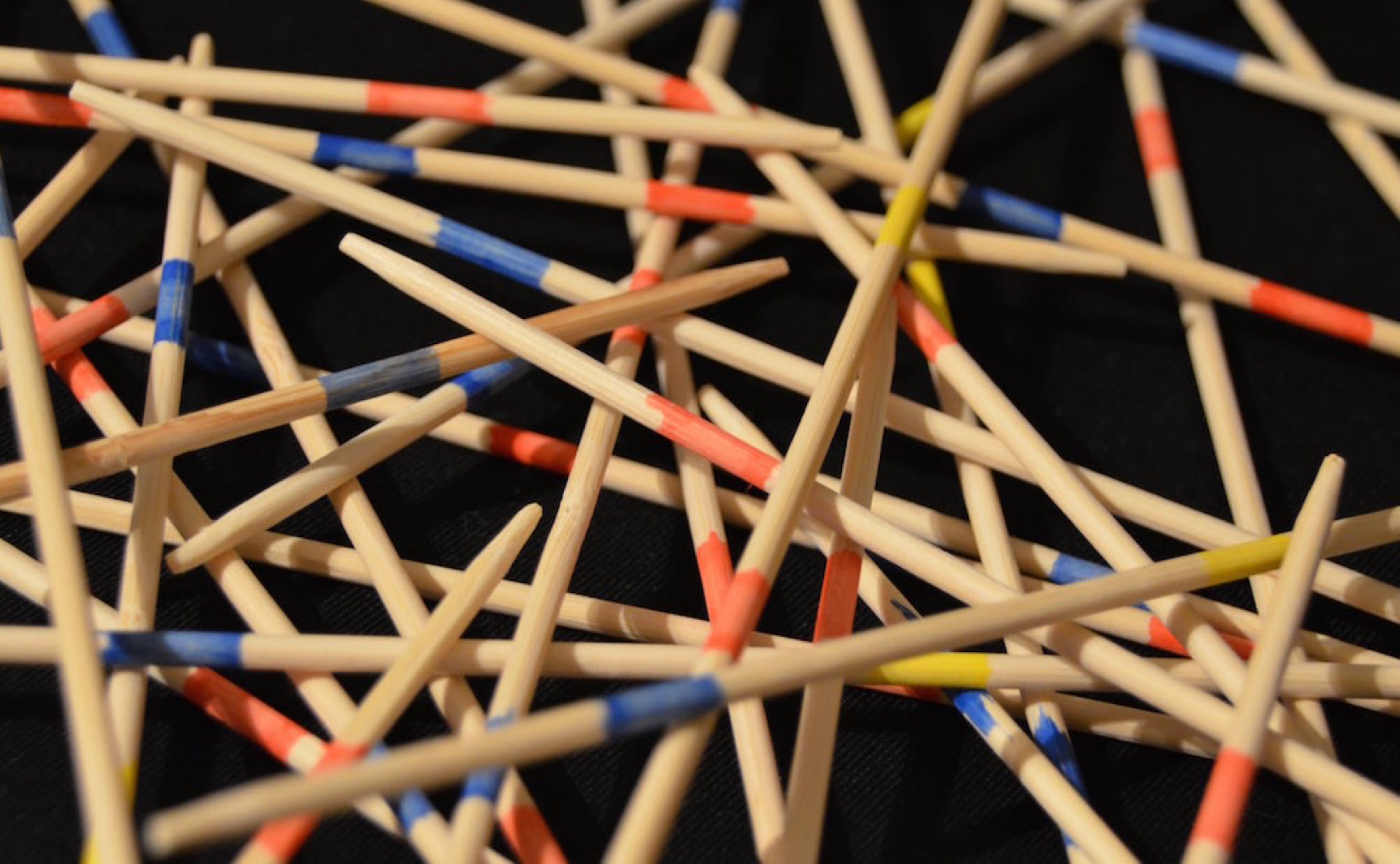}} \\
   \subfigure[]{\includegraphics[width=7.5cm]{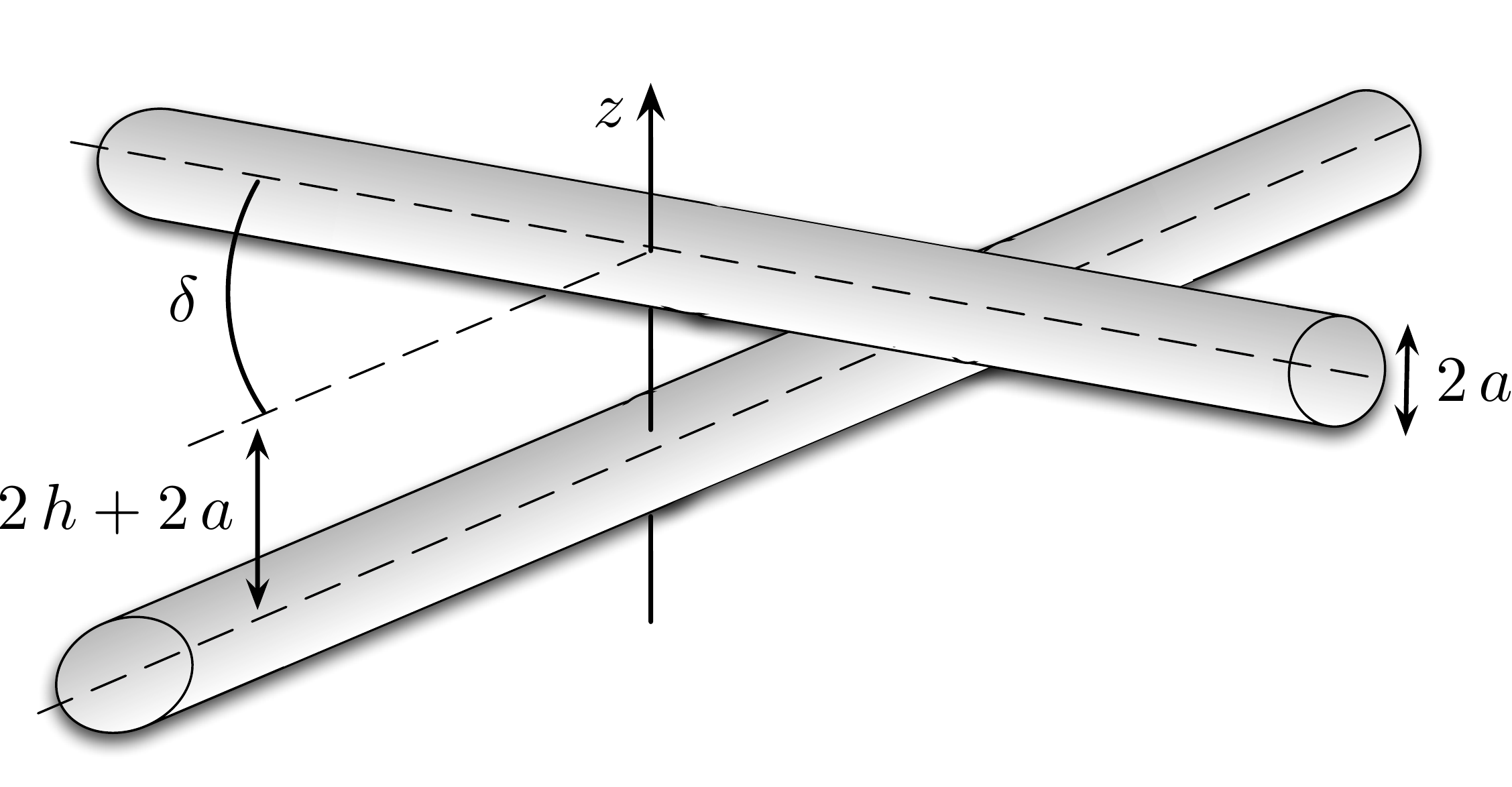}}
    \caption{(a) Representation of an array of randomly oriented rigid fibers. (b) Schematic of the system composed of two fibers of radius $2\,a$. The $z$-axis defines the position where the two fibers, having a tilt angle $\delta$, are the closest, i.e., when their axis are separated by a distance $2\,h+2\,a$.}\label{fig:setup}
\end{figure}


\section{Experimental methods}

We consider an array of randomly oriented fibers, whose typical mesh size is large compared to the drop radius, i.e. an array made of long fibers and with a large porosity, as illustrated in Fig. \ref{fig:setup}(a); the notations we will use are given in Fig. \ref{fig:setup}(b). A drop of liquid deposited on the array encounters one of four possible fiber configurations: at equilibrium the liquid can be located (i) on a single fiber, (ii) on two parallel fibers ($h \neq 0$ and  $\delta=0$), (iii) at the point of contact of two touching crossed fibers ($h=0$ and  $\delta \neq 0$), or (iv) at the point of minimum distance between two non-touching crossed fibers ($h\neq 0$ and  $\delta \neq 0$). Only the three latter configurations result in the formation of a capillary bridge and are thus of interest in the present study.

To consider the different configurations, we use a pair of identical nylon fibers tilted with an angle $\delta$ and separated by a minimum separation distance $h$. We show a schematic of the fiber configuration in Fig. \ref{fig:setup}(b). Each fiber is held horizontal and clamped at both ends, with one fiber affixed on a rotating stage (PR01, Thorlabs) with a micrometer drive that allows for the variation of the angle $\delta$ in increments of 0.1$^\circ$. The rotating stage is mounted on a linear translation stage (PT1, Thorlabs) with a micrometer drive that allows for the variation of the vertical closest distance between the fibers $h$ in increments of 5 $\mu$m. We use various fiber radii $a \in [100;\,225]\,{\rm \mu m}$ (nylon fibers from Sufix Elite) and separation distances between fibers $h \in [0,\,6\,a]$. Nylon fibers exhibit micrometer-scale roughness, but we have not observed any noticeable hysteresis with perfectly and partially wetting fluids \cite{kumar1990}. We perform systematic experiments in the three possible fiber configurations with silicone oil (5 cSt, density $\rho=918$ kg/m$^3$, surface tension $\gamma=19.7$ mN/m, puchased from Sigma-Aldrich), which is perfectly wetting on the fibers. The capillary length that describes the scale at which the gravity effects become noticeable is defined as $\ell_c = \sqrt{\gamma/(\rho\,g)}$, where $g$ is the gravitational constant and $\gamma$ is the surface tension. This length is about 1.5 mm for silicone oil. As this capillary length is usually larger than the typical height $H$ of the liquid in our experiments, we first assume that gravity effects can be neglected. We shall discuss this assumption when large volumes of fluid and/or large fibers are used, as the Bond number of the system, $Bo=\rho\,g\,{(2\,a)}^2/\gamma$, becomes larger than one.

\begin{figure*}
\centering
  \includegraphics[width=0.8\textwidth]{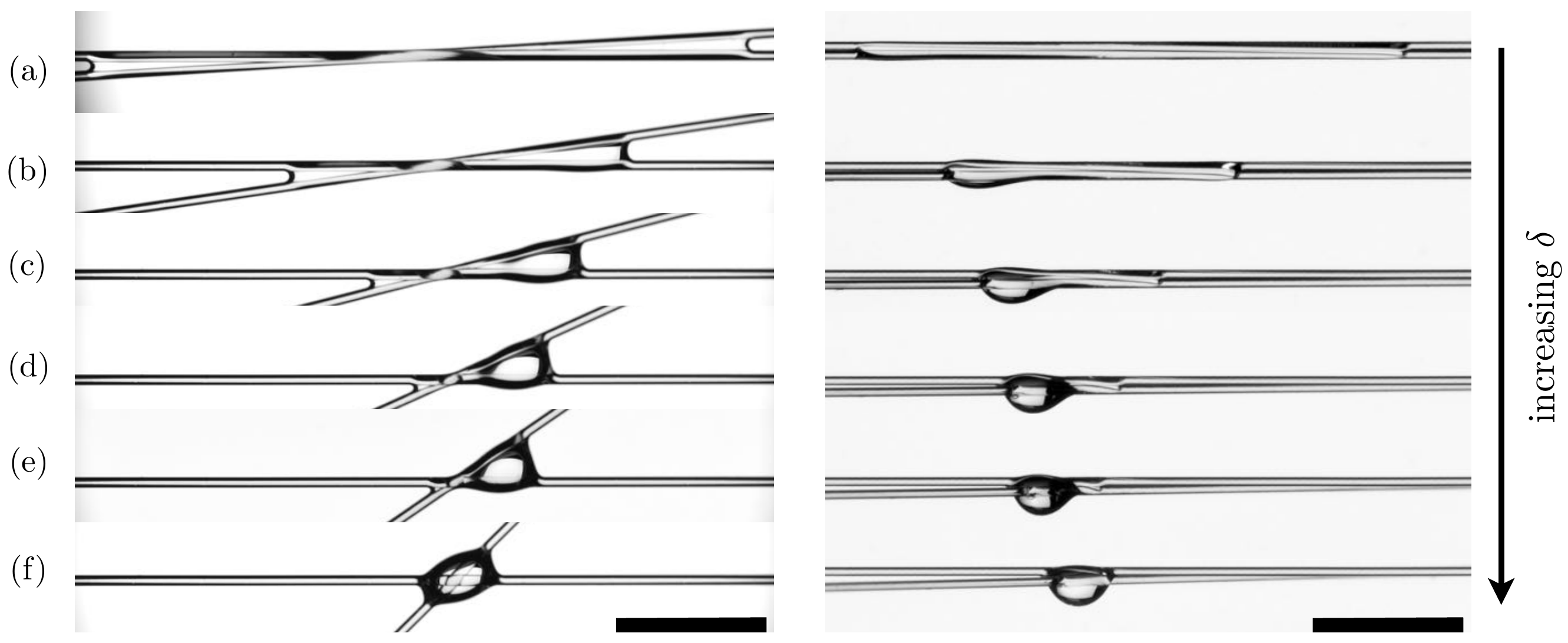}
 \caption{Evolution of the morphology of a drop of volume $V = 2\, \mu\ell$ of silicone oil (5 cSt) on two touching crossed fibers ($h=0$) of radius $a = 150 \,{\rm \mu m}$ as the angle $\delta$ between the fibers is varied: top (left) and side (right) views. The liquid starts in a column state for a small tilting angle, (a) $\delta=4^{\rm o}$. As the angle is increased, a mixed morphology is observed: (b) $\delta=9^{\rm o}$, (c) $\delta=15^{\rm o}$, (d) $\delta=25^{\rm o}$ and (e) $\delta=35^{\rm o}$. Finally, at larger angles, (f) $\delta=45^{\rm o}$, a drop lies at the crossing point of the fibers. Scale bars are 5 mm.}\label{figure2_touching}
\end{figure*}

\begin{figure*}
\centering
  \includegraphics[width=0.8\textwidth]{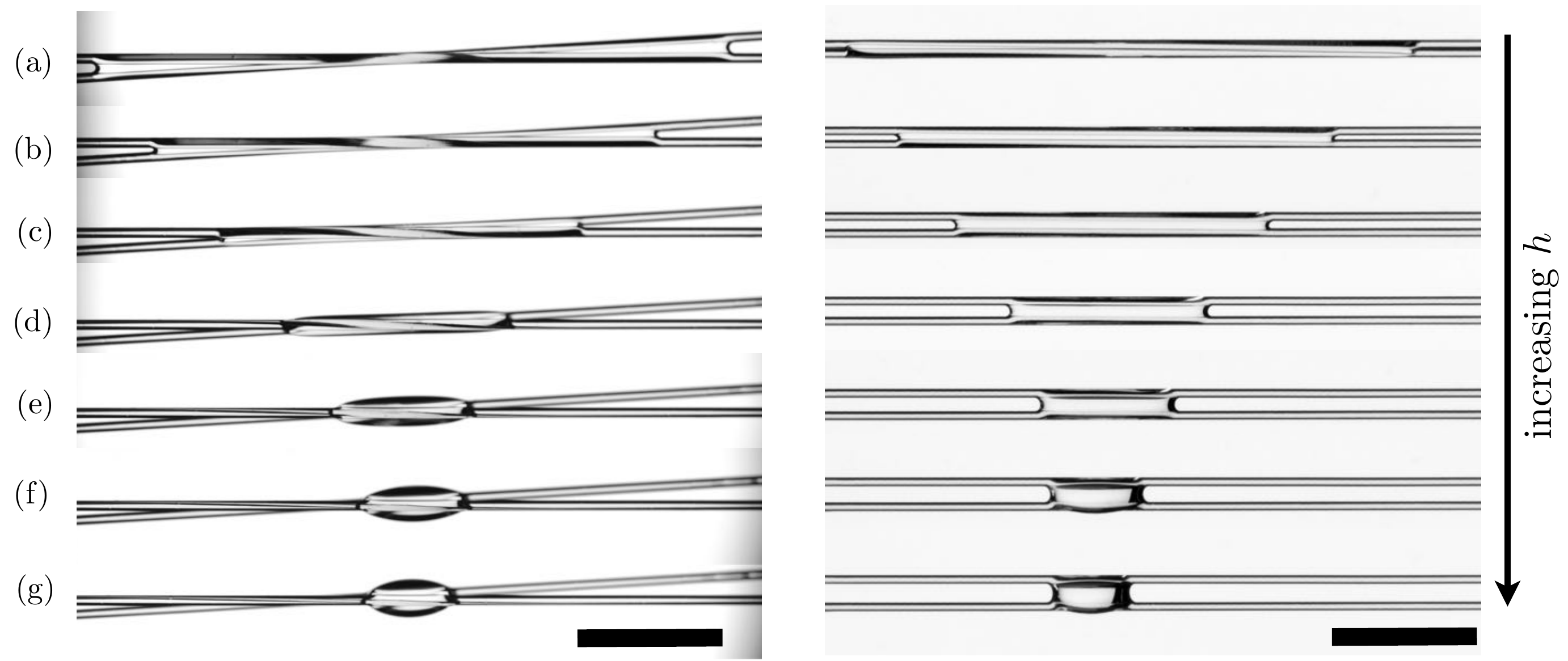}
 \caption{Evolution of the morphology of a drop of volume $V = 2\, \mu\ell$ of silicone oil (5 cSt) on two non-touching crossed fibers of radius $a = 150 \,{\rm \mu m}$, tilted by an angle $\delta=4^{\rm o}$ as the distance $h$ between the fibers is increased: top (left) and side (right) views. The liquid starts in a column state at small separation distances, (a) $h=0$. As the distance $h$ is increased, the liquid adopts a drop morphology (f)-(g). Scale bars are 5 mm.}\label{figure3}
\end{figure*}

In a typical experiment, we dispense a known volume $V \in [0.5;\,8]\,{\rm \mu \ell}$ of liquid using a micropipette (Eppendorf) on a pair of crossed fibers separated vertically by a separation distance $h$. We increase the angle between the fibers incrementally until $\delta \simeq 90^\circ$ and then decrease $\delta$ incrementally. For each step in $\delta$, the equilibrium state of the liquid is captured from the top and the side views with cameras (Nikon cameras D5100 and D7100 and $105$ mm macro objectives) as illustrated in Figs. \ref{figure2_touching} and \ref{figure3}. The top view allows the measurement of the angle between the fibers and wetting length, while the side view permits the measurement of the separation distance between the fibers and the discrimination between the different states. Three possible liquid morphologies are observed: the drop state, the mixed morphology and the column state. In the drop state, the liquid collects in a single drop centered on the point where the distance between the fibers is the smallest, i.e. the node. The column morphology corresponds to the spreading of the liquid along the fibers. In this morphology, the height of the liquid remains of the same order of magnitude as the separation distance between the fibers. Finally, the mixed morphology is defined by the coexistence of a column and a drop lying at one end of a column. The position of the drop, i.e. the side of the column where it is located, is random and due to external noise when changing the tilting angle or the inter-fiber separation.

Experimentally, as the angle between the fibers is increased (Fig. \ref{figure2_touching}), the length of the column of liquid decreases and the liquid switches to a mixed morphology. As the angle between the fibers is further increased, the liquid configuration becomes a drop. Then, when decreasing the angle between the fibers, the drop reverts back to the mixed morphology and eventually elongates into a column. We can also keep the tilt angle $\delta$ constant and increase the separation distance $h$ (Fig. \ref{figure3}).
For the particular case of parallel fibers ($\delta=0$), we increase the separation distance $h$ incrementally until the drop morphology is observed. The same procedure is followed to measure the separation distance and to observe the liquid morphology. The change of morphology between the column and the drop state can be observed in the plane defined by the two fibers. Indeed, as $h$ increases, the transition occurs when the liquid overspills the fibers. At the transition, the liquid collects in a drop. For instance, the morphology can be discriminated between figures \ref{figure3}(e) and \ref{figure3}(f).


\section{Analytical modeling}

\begin{figure}
    \centering
   \subfigure[]{\includegraphics[width=7cm]{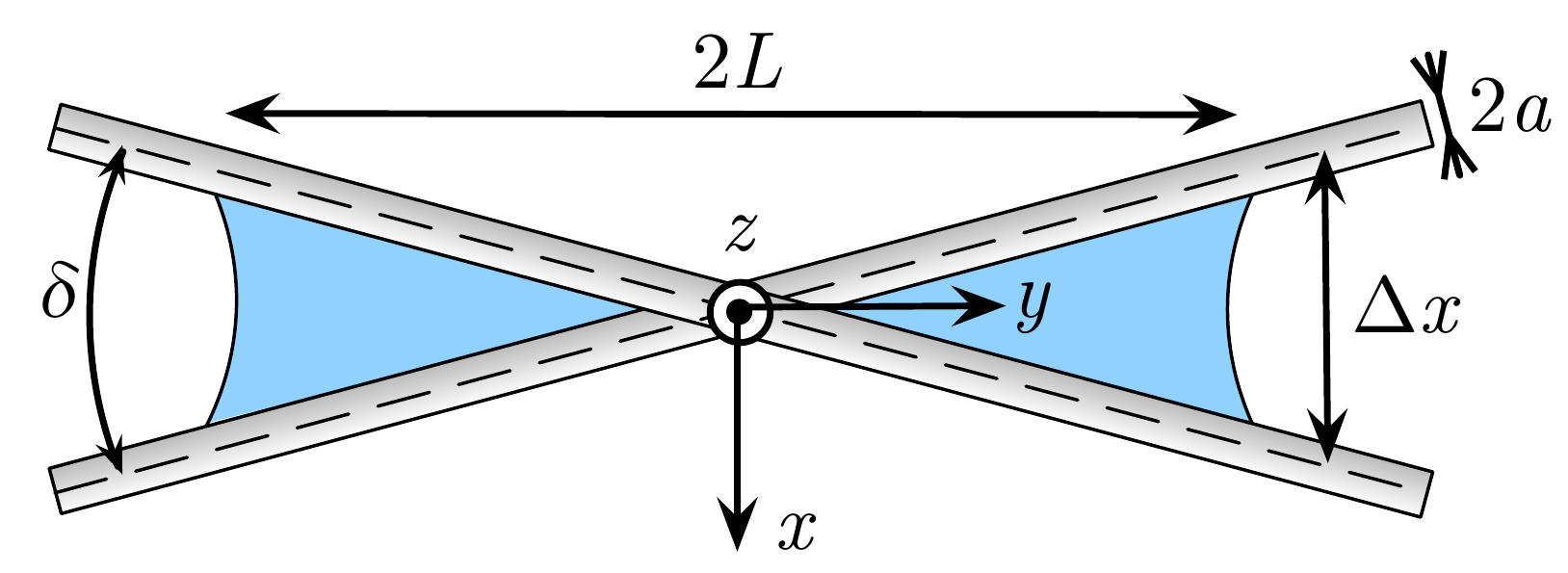}}
   \subfigure[]{\includegraphics[width=4.5cm]{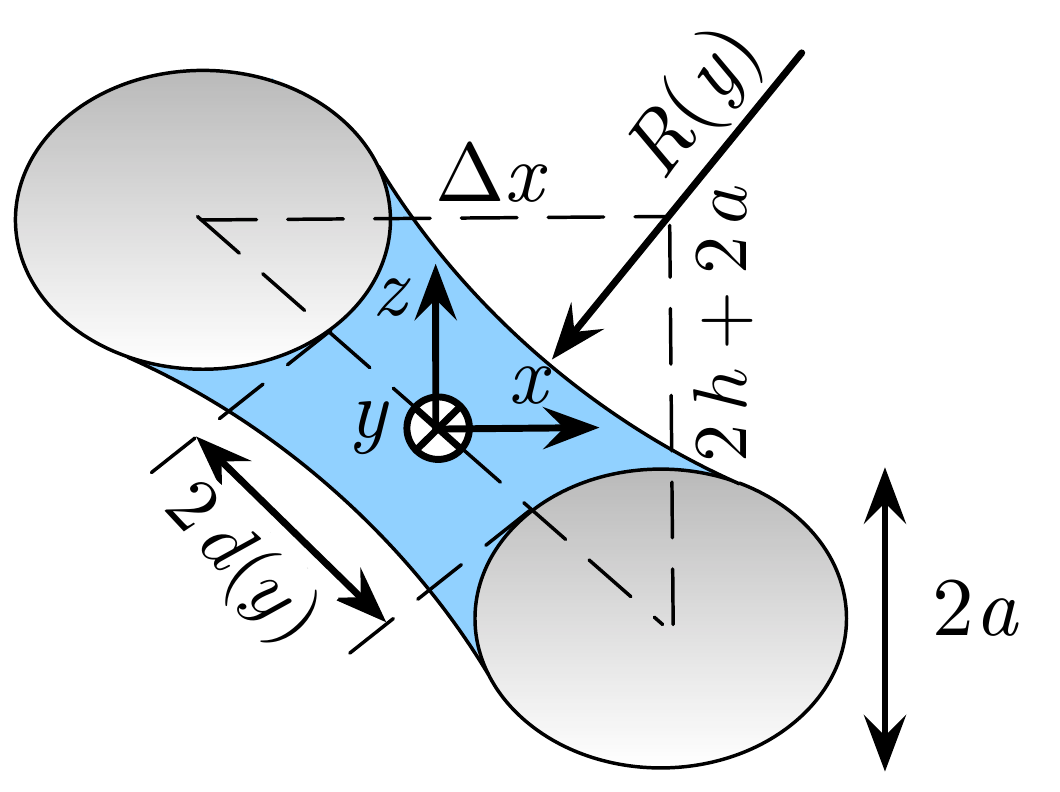}}
    \caption{Schematic and notations of the system composed of two fibers of radius $2\,a$. (a) Top view and (b) cross-section view.}\label{schema}
\end{figure}

We consider two rigid fibers having a cylindrical cross-section, separated by a minimum distance $h$ and tilted by an angle $\delta$. We can define a system of coordinates $Oxyz$ as represented schematically in Fig. \ref{schema}. When needed, we use the fiber radius $a$ to construct dimensionless parameters: the dimensionless inter-fiber distance $\tilde{d}=d/a$, the dimensionless spacing distance $\tilde{h}=h/a$, the dimensionless wetting length $\tilde{L}=L/a$, the dimensionless cross-sectional area $\tilde{A}=A/a^2$ and the dimensionless volume $\tilde{V}=V/a^3$. Provided that the fibers are not parallel, i.e. that the tilting angle is not zero, a drop of liquid lying on two fibers will travel towards the point where the fibers are the closest, which we refer to as the ``kissing point'' if the fibers are in contact.

We consider a liquid that has a contact angle $\theta_E$ and is in a column state on a pair of fibers characterized by ($\delta$, $h$, $a$). This morphology consists of a long column of liquid, with varying cross-section and a constant height. The shape of the surface of the cross-section of the column is defined by its dimensionless radius of curvature $\tilde{R}=R/a$ and the angle between the line connecting the centers of the fibers and the radius to the liquid-fiber-air boundary $\alpha$ [Fig. \ref{schema}(b)]. We define the equilibrium configuration of the general situation of two fibers that are not in contact, i.e. separated by a minimum distance $h > 0$ and tilted with an angle $\delta > 0$.

The inter-fiber distance, $2\,d(y)$, varies as a function of the distance $y$ to the point $O$ where the two fibers are the closest [Fig. \ref{schema}(a)]. $\Delta x$ is the distance between the axes of the two fibers projected in the plane ($x\,y$), and $2\,h+2\,a$ is the closest distance between the axes of the two fibers. We have
  \begin{equation}
\left(\Delta x\right)^2+4\,(a+h)^2=4\,\left[d(y)+a\right]^2 \quad \text{and} \quad \tan\left(\frac{\delta}{2}\right)=\frac{\Delta x}{2\,y}.
\end{equation}
Using these two expressions, we obtain
\begin{equation}\label{toto1}
\tilde{d}(y) = \frac{d(y)}{a}=\left[\tilde{y}^2\,\tan^2\left(\frac{\delta}{2}\right)+(1+\tilde{h})^2\right]^{1/2}-1,
\end{equation}
where $\tilde{y} = y/a$.

We observe that for $\delta=0$ and $\tilde{h} >0$, we recover the expression derived by Princen for parallel fibers \cite{Princen1970}, whereas for $\tilde{h}=0$ and $\delta>0$ we obtain the situation of touching crossed fibers \cite{Sauret2014}. Using geometrical arguments, we define the radius of curvature $\tilde{R}$,
\begin{equation} \label{3}
     \tilde{R} = \frac{R}{a}=\frac{1+\tilde{d}-\cos\alpha}{\cos(\alpha+\theta_E)},
\end{equation}
and the liquid cross-sectional area,

\begin{eqnarray}\label{2}
    \tilde{A} = \frac{A}{a^2} & = & \tilde{R}^2\,\Bigl[2\,\alpha+2\,\theta_E-\pi+\sin[2\,(\alpha+\theta_E)]\Bigr] \nonumber\\
                                  & & +4\,\tilde{R}\,\sin\alpha\,\cos(\alpha+\theta_E)-2\alpha+\sin(2\,\alpha).
\end{eqnarray}

To determine analytically the cross-sectional shape of the column morphology at the equilibrium, we assume that at each distance $\tilde{y}$ from the ``kissing'' point, the cross-sectional shape only depends on the distance between the two fibers, $\tilde{d}(\tilde{y})$ and on the contact angle $\theta_E$. A force balance on an infinitesimal volume $\text{d}V=A\,\text{d}L$ leads to the equilibrium condition \cite{Sauret2014}:
\begin{equation}\label{1}
    4\,\left[\left(\frac{\pi}{2}-\alpha-\theta_E\right)\,\tilde{R}-\alpha\,\cos \theta_E\right]+\frac{\tilde{A}}{\tilde{R}}=0.
\end{equation}

Substituting the expression of $\tilde{A}$ given by the relation (\ref{2}) in equation (\ref{1}) leads to a quadratic equation for the radius of curvature $\tilde{R}$ \cite{Princen1970}:
\begin{eqnarray} \label{Equation_R}
    & & \tilde{R}^2\,\Bigl[\pi-2\,\alpha-2\,\theta_E+\sin[2\,(\alpha+\theta_E)]\Bigr] -2\alpha+\sin(2\,\alpha) \nonumber\\
& & +4\,\tilde{R}\,\Bigl[\sin\alpha\,\cos(\alpha+\theta_E)-\alpha\,\cos\theta_E\Bigr] = 0.
\end{eqnarray}

Therefore, for a given liquid, i.e. a specified value of the contact angle $\theta_E$, equation (\ref{Equation_R}) can be solved to obtain the dimensionless radius of curvature $\tilde{R}$ as a function of $\alpha$.

For the particular case of a perfectly wetting liquid ($\theta_E=0$), we obtain a simple expression of $\tilde{R}$ using equation (\ref{Equation_R}) \cite{Protiere2012,Sauret2014}. Substituting this expression in equation (\ref{3}) and with equations (\ref{toto1}), we obtain a direct relation between $\tilde{y}$ and $\alpha$:
\begin{eqnarray}\label{5}
    & & \sqrt{\tilde{y}^2\,\tan^2\left(\frac{\delta}{2}\right)+\left(1+\tilde{h}\right)^2} \qquad\qquad \nonumber \\
& & \qquad  =\left(1+\left[\sqrt{\frac{\pi}{2\,\alpha-\sin(2\,\alpha)}}-1\right]^{-1}\right)\,\cos\alpha. 
\end{eqnarray}

For $\delta=0$, we recover from equation (\ref{5}) the expression derived by Princen \cite{Princen1970} and Proti\`ere et al. \cite{Protiere2012} to describe the liquid morphology on parallel fibers. For $\tilde{h}=0$, we obtain the expression derived by Sauret et al. for touching crossed fibers \cite{Sauret2014}.

To determine the maximum volume of fluid that can be contained in a column morphology that is a symmetric state, we consider the expression of the volume of liquid lying on the fibers in this morphology:
\begin{eqnarray} \label{numero1}
    \tilde{V} & = & \int_{-\tilde{L}}^{\tilde{L}}\,\tilde{A}(\tilde{y})\,\text{d}\tilde{y},
\end{eqnarray}
where $\tilde{A}(\tilde{y})$ is the cross-sectional area and $\tilde{L}$ the half-length of the column. Imposing a constraint on the volume of liquid $\tilde{V}$ leads to a unique value of the wetted length $2\,\tilde{L}$.

In addition, solving the quadratic equation (\ref{Equation_R}) for $\tilde{R}$ and substituting the solution in relation (\ref{3}), we observe that $\tilde{d}$ reaches a maximum value $\tilde{d}_{max}$ when varying $\alpha$ for a given $\theta_E$. Therefore, if the local inter-fiber distance $\tilde{d}(\tilde{y})$ is larger than $\tilde{d}_{max}$, the column state cannot exist. This condition defines the maximum length of a liquid column state, since this corresponds to $\tilde{d}(\tilde{L}_{max})=\tilde{d}_{max}$. Using relation (\ref{toto1}), we write the maximum spreading length:
\begin{equation}
    \tilde{L}_{max}=\frac{\left[(1+\tilde{d}_{max})^2-(1+\tilde{h})^2\right]^{1/2}}{\tan(\delta/2)}.
\end{equation}

We observe that the wetting length increases when decreasing the tilting angle $\delta$ and the separation distance $\tilde{h}$. Note in particular that for touching crossed fibers ($\tilde{h}=0$) and a perfectly wetting liquid, we have $\tilde d_{max}=\sqrt{2}$.

For a given separation distance $\tilde h$ and tilt angle $\delta$, the maximum wetting length defines the maximum volume of the column and thus the regime of existence of the column state. Indeed, the maximum volume of liquid that can be at equilibrium in a column state is defined by
\begin{eqnarray} \label{V_max}
    \tilde{V}_{max} & = & \int_{-\tilde{L}_{max}}^{\tilde{L}_{max}}\,\tilde{A}(\tilde{y})\,\text{d}\tilde{y}.
\end{eqnarray}

For a volume of liquid $\tilde V$ larger than $\tilde{V}_{max}$, the liquid would not be able to spread in a column state and could be either in a mixed morphology or in a drop state as we shall see in the following. However, even for a volume of liquid $\tilde V<\tilde V_{max}$, we need to compare the surface energies of all possible morphologies defined as $E=\gamma\,A_{LV}-\gamma\,\cos\theta_E\,A_{SL}$ where $A_{LV}$ and $A_{SL}$ are the liquid-air and liquid-fiber surface areas. These energies are minimized to determine which morphology will be preferentially adopted by the liquid.

The transition between the drop state and the mixed morphology is more complex as the shape of the drop between two fibers does not have an analytical description. We assume that the surface energy associated with the drop morphology is that of a sphere of equivalent radius $[3\,\tilde{V}/(4\,\pi)]^{1/3}$, pierced by two fibers. Note that Proti\`ere et al. \cite{Protiere2012} show that a better quantitative agreement between theoretical and experimental results can be obtained by modeling the drop with a shape close to a hemisphere with an energy equal to :
\begin{equation} \label{E_drop}
\tilde{E}_{drop}=\frac{{E}_{drop}}{\gamma\,a^2}=0.6\,\left[(36\,\pi)^{1/3}\,\tilde{V}^{2/3}-\pi\,\sqrt{\left(6\,\tilde{V}/\pi\right)^{2/3}-4\,\tilde{d}^2}\right],
\end{equation}
where the pre-factor $0.6$ is empirical and takes into account that the shape of the liquid in the drop state is not exactly a sphere. The corresponding surface energy associated with the column morphology is
\begin{equation}\label{Ecolumn}
    \tilde{E}_{col}=\frac{{E}_{col}}{\gamma\,a^2}=4\,\int_{-\tilde{L}}^{\tilde{L}}\left[\tilde{R}\left(\frac{\pi}{2}-\theta_E-\alpha\right)-\alpha\,\cos\theta_E\right]\,\text{d}\tilde{y}
\end{equation}
Note that this formulation allows us  to recover the expression previously obtained for two parallel fibers\cite{Protiere2012} as in this situation $\alpha$ and $\tilde{R}$ are constant along the column and the energy reduces to
\begin{equation} \label{E_col}
   \tilde{E}_{col}=8\,\tilde{L}\,\left[\,\left(\frac{\pi}{2}-\theta_E\,-\alpha\right)\,\tilde{R}-\alpha\,\cos\theta_E\right].
\end{equation}

In the present situation, we evaluate equation (\ref{Ecolumn}) numerically using the expressions for $\alpha$ and $\tilde R$ obtained in the previous section for varying distance $\tilde{d}(y)$ between the fibers similarly to the derivation by Sauret et al. \cite{Sauret2014}. The drop shape on a pair of fibers is much more complex to describe as there is no analytical expression that captures the shape of the drop.

In addition, we also need to impose a constraint on the volume: the liquid can be either in a column morphology $\tilde{V}_{col}$, in a drop morphology $\tilde{V}_{drop}$ or in a mixed state but the total volume of liquid $\tilde V$ should always satisfy $\tilde{V}=\tilde{V}_{col} + \tilde{V}_{drop}$. The dimensionless energy $\tilde E$ of the system $\tilde{E}=\tilde{E}_{col} + \tilde{E}_{drop}$ reaches a minimum for a given volume $\tilde V$, a given tilt angle $\delta$ and a given separation distance $\tilde h$. In addition, we assume that there is no activation barrier between the various morphologies. By doing so, we observe qualitatively the transition between the mixed morphology and the drop state but no quantitative evolution can be obtained. Therefore, this transition is captured experimentally only.


\section{Morphology diagrams}

\subsection{Parallel fibers}

\begin{figure}
    \centering
\includegraphics[width=7.5cm]{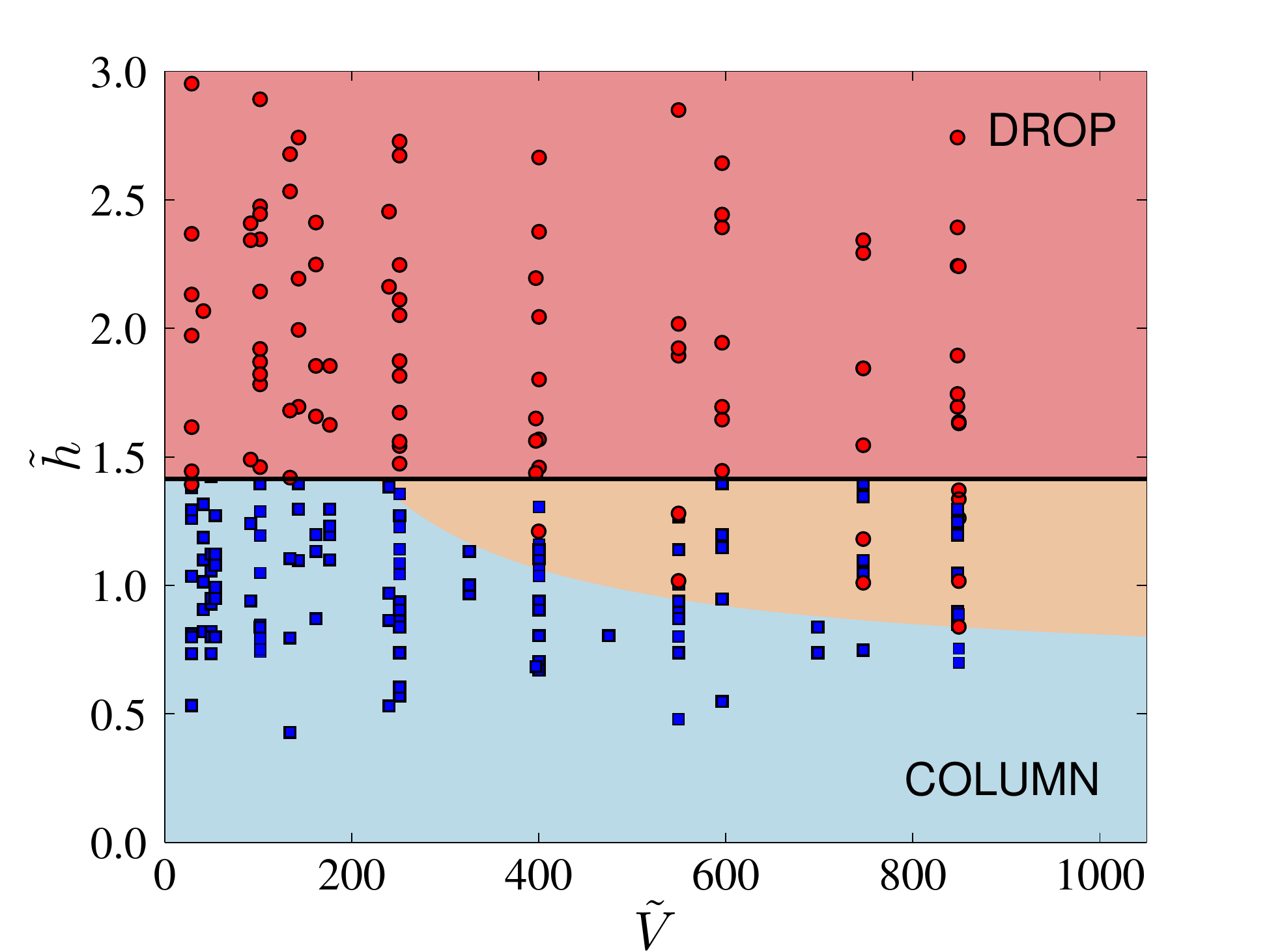}
\caption{Morphology diagram in the parameter space ($\tilde{V}$, $\tilde{h}$) using silicone oil ($\theta_E=0\, ^\circ$) on parallel fibers ($\delta=0\, ^\circ$) of radii $a=[100,\,150,\,230]\,\mu{\rm m}$ and a volume of liquid $V \in [0.5,\,4]\,\mu\ell$. Red circles show the drop morphology and blue squares the column morphology. The light orange region corresponds to the region where both morphologies are observed. The horizontal solid line corresponds to the theoretical maximum separation distance where the column state is possible, $\tilde{d}_{max}=\sqrt{2}$.}\label{fig:parallel}
\end{figure}

We first conducted experiments with drops of silicone oil on parallel nylon fibers to verify the analytical model for perfectly wetting liquids. The experimental results are reported in Fig. \ref{fig:parallel} in a morphology diagram of $\tilde{h}$ as a function of $\tilde{V}$ (the results for dodecane, i.e, a partially wetting liquid are reported in the appendix). We find that the drop-column transition occurs for the maximum separation $\tilde{h}=\tilde{d}_{max}=\sqrt{2}$, which is in agreement with the analytical solution derived in the previous section and consistent with previous experimental results obtained for this geometry. For larger volumes of liquid, $\tilde{V} > 400$, we observe a coexistence region in which for a given $\tilde h$ and $\tilde{V}$ the liquid can either be in a drop state or a column morphology. This coexistence region widens with increasing volumes. The coexistence region was also observed by Proti\`{e}re et al. \cite{Protiere2012}

We can understand the coexistence region by considering the Bond number of the system, defined as $Bo=\rho\,g\,{(2\,a)}^2/\gamma$, where $g$ is the gravitational constant, $2\,a$ is a characteristic length scale associated to the separation distance and $\gamma$ is the surface tension. The Bond number describes the relative influence of the gravitational force relative to surface tension effects. Within the coexistence region, we generally find $Bo > 1$, which is an indication that the effects of gravity on the liquid cannot be neglected. Gravity can hinder the liquid from spreading into the more stable configuration, i.e. the column state, which results in the coexistence region observed. However, capturing quantitatively this transition would require numerical simulations to define the shape of a drop and a column in the presence of gravity, which is out of the scope of the present study. Our studies have been performed in an horizontal plane, but we can infer that gravitational effects on the parameters space of the drop could be modified slightly when this plane is tilted.

\subsection{Touching crossed fibers}

\begin{figure}
    \centering
\includegraphics[width=7.5cm]{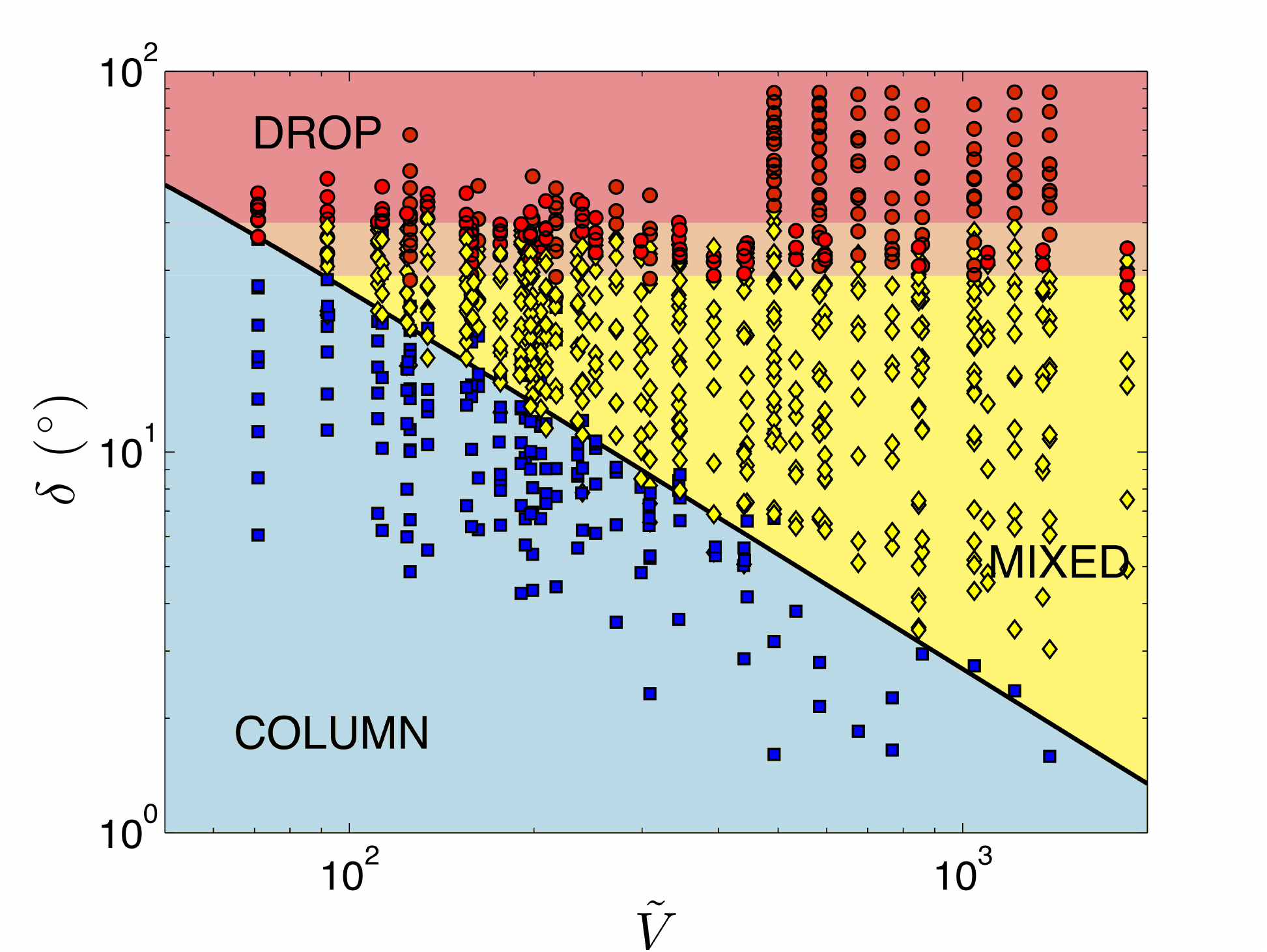}
\caption{Morphology diagram for touching crossed fibers of radii $a=[100,\,150,\,180,\,230]\,\mu{\rm m}$ in the parameter space ($\tilde{V}$, $\delta$) using a volume $V\in [0.5,\,7]\,\mu\ell$ of silicone oil ($\theta_E=0\, ^\circ$). Red circles show the drop morphology, blue squares the column morphology and yellow diamonds are the mixed morphology. The light orange region corresponds to the parameters where both drop and the mixed morphology are observed. The solid line corresponds to the theoretical maximum angle, for a given volume $\tilde{V}$, below which the column state is possible.}\label{fig:crossed}
\end{figure}

We then perform experiments with two touching crossed-fibers ($\tilde{h}=0$ and $\delta > 0$) to further explore the validity of the analytical model  for liquids with zero contact angle. We report the observed morphologies in Fig. \ref{fig:crossed}. The column morphology is observed at small enough tilt angle $\delta$ or rescaled volume $\tilde{V}$. On the morphology diagram, we also plot the analytical prediction for the transition between the column state and the two other morphologies, which represents the volume $V_{max}$ beyond which a column state cannot exist between crossed fibers (see equation (\ref{V_max})). We observe a good agreement between the analytical prediction and the experimental results.

The transition between the mixed morphology and drop state is not captured by an analytical expression as the exact shape of the drop is not known and because of the rough estimate of equation (\ref{E_drop}). Our experimental results, however, indicate a coexistence region (light orange in the diagram) where both the mixed and the drop morphologies are present. The transitions from the mixed morphology to the coexistence region and from the coexistence region to the drop state appear to be independent of the $\tilde{V}$ for $\tilde{V} > 100$. We can estimate the Bond number at which the transition between the mixed and drop states occurs. The results suggest that the transition between the mixed and drop morphologies is also independent of the Bond number. It is possible, however, that the transition is a weak function of $\tilde{V}$ and the Bond number, but we are unable to detect it through our experiments.

\subsection{Separated crossed fibers}

\begin{figure}
    \centering
\includegraphics[width=7.5cm]{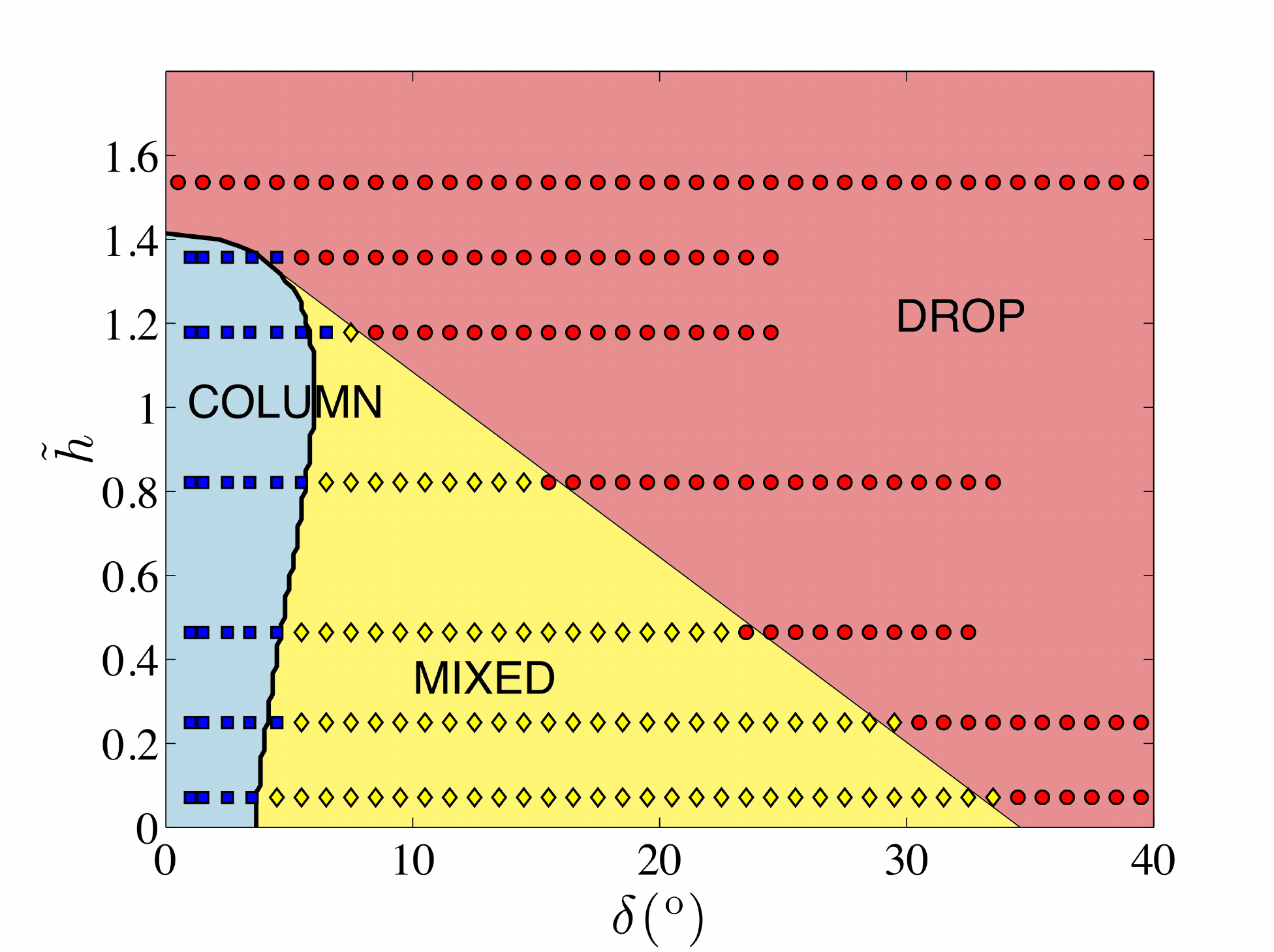}
\caption{Morphology diagram for a drop of silicone oil places on separated crossed fibers of radii $a=150\,\mu{\rm m}$ in the parameter space ($\delta$, $\tilde{h}$) using a volume $\tilde{V}=592$ of silicone oil ($\theta_E=0\, ^\circ$). Red circles show the drop morphology, blue squares the column morphology and yellow diamonds are the mixed morphology. The thick solid line is the prediction of our model. }\label{fig:separated}
\end{figure}

In this last configuration, the tilt angle $\delta$ and the minimum spacing distance $\tilde h$ are both non-zero and influence the resulting morphology. In addition, the dimensionless volume $\tilde{V}$ is a parameter to consider, which leads to a huge parameter space to investigate. To compare experimental measurements with our analytical model, we performed experiments with drops of silicone oil at constant volume, $\tilde{V}=592$ (corresponding to $2\,\mu\ell$ on fibers of radii $a=150\,\mu{\rm m}$) and vary the tilt angle as well as the separation distance. The results of our investigation is presented in Fig. \ref{fig:separated}. Similar to the touching crossed-fiber system we observe three possible morphologies: drop, mixed and column, each defined in a region of our parameter space ($\delta$, $\tilde{h}$). We also report our analytical prediction in this diagram (black thick line) that captures the region where the column state is observed. We should emphasize that the horizontal axis ($\tilde{h}=0$) corresponds to the situation of touching crossed fibers and we again observe the transition from a column morphology to a mixed morphology and eventually a drop at large angles. The vertical axis ($\delta=0$) corresponds to a pair of parallel fibers, and in agreement with our previous results (Fig. \ref{fig:parallel}) we only observe two states: column and drop. Fig. \ref{fig:separated} further confirms the possibility to predict the observed morphology on two fibers randomly oriented in space.

\section{Conclusions}

\begin{figure}
    \centering
\includegraphics[width=8.5cm]{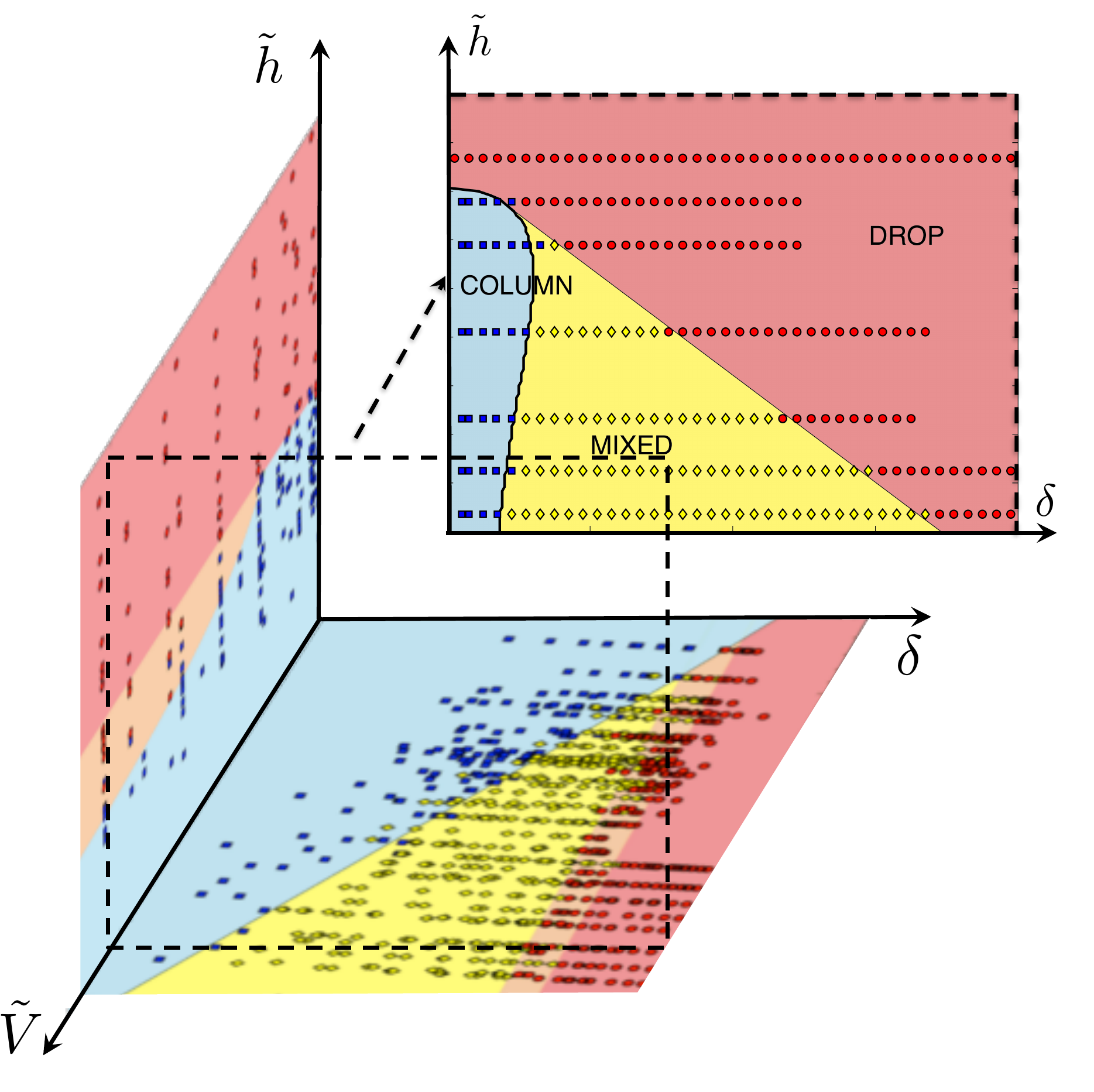}
\caption{Morphology diagram for a pair of fibers randomly oriented wetted by a drop of silicone oil ($\theta_E=0\, ^\circ$) in the parameter space ($\tilde{V}$, $\delta$, $\tilde{h}$). Red circles show the drop morphology, blue squares the column morphology and yellow diamonds are the mixed morphology. The light orange region corresponds to the region where both morphologies are observed.}\label{fig:3D}
\end{figure}

In this paper, we have investigated experimentally and theoretically the wetting morphologies on a pair of randomly placed and oriented fibers. In agreement with previous studies, we show that in the most general case three morphologies are observed: a column morphology, a single drop located at the node, and a mixed morphology with one drop at one end of a column. We report our analytical and experimental findings in a new three-dimensional morphology diagram shown in Fig. \ref{fig:3D}, which captures all possible situations. The three relevant parameters to describe the wetting morphologies between two randomly oriented fibers are the volume of liquid $\tilde{V}=V/a^3$, the tilt angle $\delta$ and the minimum separation distance between fibers $\tilde{h}=h/a$. Additionally, we show that the results obtained previously for the more restrictive cases of parallel and touching crossed fibers can be recovered for zero spacing distance between the fibers coupled, respectively, with a zero and non-zero angle between the fibers. Thus, the three possible situations that we have highlighted in this article allow us to define a full three-dimensional diagram ($\delta$, $\tilde{h}$, $\tilde{V}$) to predict the liquid morphology on a pair of randomly oriented fibers (Fig. \ref{fig:3D}).

The analytical model that describes the transition between the different morphologies of the liquid on non-touching fibers is validated by experimental results. It should be noted that the model is applicable to perfectly and partially wetting liquids. The characterization of the transitions between the three different wetting morphologies will help to better understand the behavior of a wet fiber array and describe the capillary interactions generated by a liquid bridge between two fibers. This approach is especially relevant to systems composed of flexible fibers where the capillary force can lead to the clustering of fibers \cite{Bico2004,Duprat2012}.

Our results suggest that a system of fibers can be used to manipulate liquids on a micro scale. For example, by mechanically altering the angle and/or spacing distance between the fibers, or by triggering changes in the liquid volume through condensation or evaporation,\cite{Duprat2013,Boulogne2015} we can change the morphology adopted by the liquid on the fibers. Additionally, since the model that we have proposed is applicable to both perfectly and partially wetting liquids, we can consider using the transitions between wetting morphologies on fibers to estimate the contact angle of liquids on fibers. Presently, the contact angle of a liquid on fibers can be computed using a method proposed by Carroll \cite{Carroll1976} that involves solving elliptic integrals. A possible alternative to such a cumbersome method, for example, would be to use the transition between the drop and column states of a liquid on parallel fibers.

\section*{Acknowledgements}
We thank H\'el\`ene Lannibois-Drean and François Vianney from Saint-Gobain Research and Pierre-Brice Bintein for providing the SEM pictures of glass wool. FB acknowledges that the research leading to these results partially received funding from the People Programme (Marie Curie Actions) of the European Union's Seventh Framework Programme (FP7/2007-2013) under REA grant agreement 623541. ED is supported by set-up funds by the NYU Polytechnic School of Engineering. HAS thanks the Princeton MRSEC for partial support of this research.

\appendix

\section{Partially wetting liquid}

Most experiments performed to investigate the wetting morphologies on a pair of fibers use silicone oil that is a perfectly wetting liquid ($\theta_E=0^{\rm o}$). To ensure that our model correctly captures the influence of the contact angle, we also performed experiments using dodecane (density $\rho=748$ kg/m$^3$, surface tension $\gamma=25.4$ mN/m, purchased from Sigma-Aldrich), which is a partially wetting liquid and compared these results to the analytical prediction. The  contact angle of the dodecane on nylon fibers has been measured and estimated to be $ 13 \pm 1 \, ^\circ$. The liquid-fiber contact angle was measured using the shape of a drop on a single fiber \cite{Carroll1976}.

\medskip

\begin{figure} [h!]
    \centering
\includegraphics[width=7.5cm]{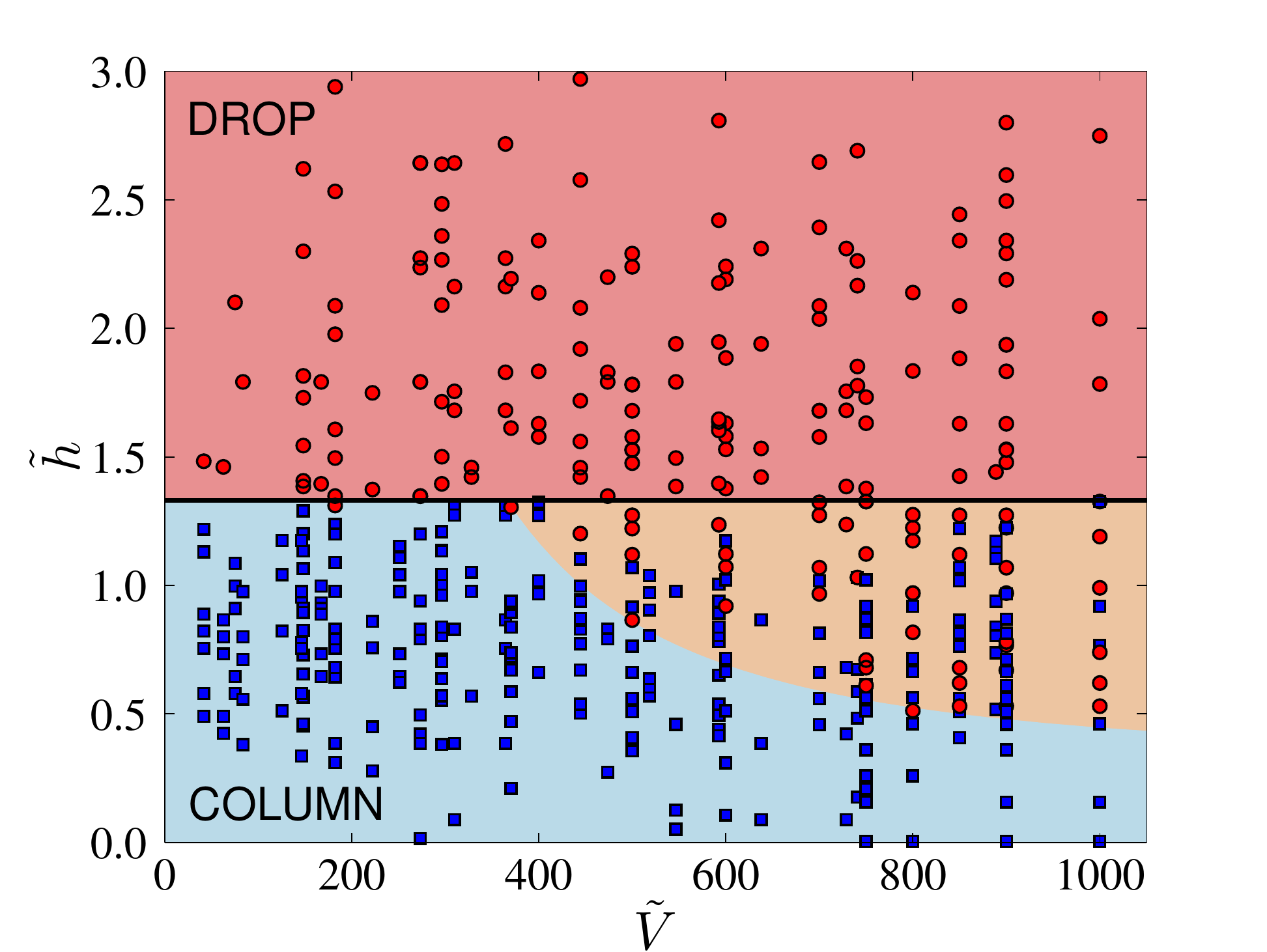}
\caption{Morphology diagram in the parameter space ($\tilde{V}$, $\tilde{h}$) using a partially wetting liquid (dodecane) on parallel fibers of radii $a=[100,\,150,\,230]\,\mu{\rm m}$ and a volume of liquid $V \in [0.5,\,4]\,\mu\ell$. Red circles show the drop morphologies and blue squares the column morphology. The light orange region corresponds to the region where both morphology are observed. The horizontal solid line corresponds to the analytical predictions for dodecane: $\tilde{h}_{max}\simeq 1.33$.}\label{fig:Appendix_parallel}
\end{figure}

We perform systematic experiments using a pair of parallel fibers varying the volume of liquid $\tilde{V}=V/a^3$ and the minimum spacing distance $\tilde{h}=h/a$. The resulting morphology diagram is shown in Fig. \ref{fig:Appendix_parallel}. The transition between the drop state and the column morphology is captured by the analytical calculation, which predicts a maximum separation $\tilde{h}_{max} \simeq 1.33$ for a liquid with contact angle of $13^{\rm o}$. As seen in the morphology diagram, for small rescaled volumes, $\tilde{V} \leq  400$, there is a sharp transition between drop and column states at $\tilde{h}_{max} = 1.33$. As with silicone oil, we observe a coexistence region (in orange) in which generally $Bo > 1$, an indication that gravity cannot be neglected in such cases. We note that the coexistence region observed for dodecane is larger than that for silicone oil, suggesting that the larger contact angle and the hysteretic effect in the contact line on nylon fibers makes it more difficult to spread on the fibers.

\begin{figure} [h!]
    \centering
   \includegraphics[width=7.5cm]{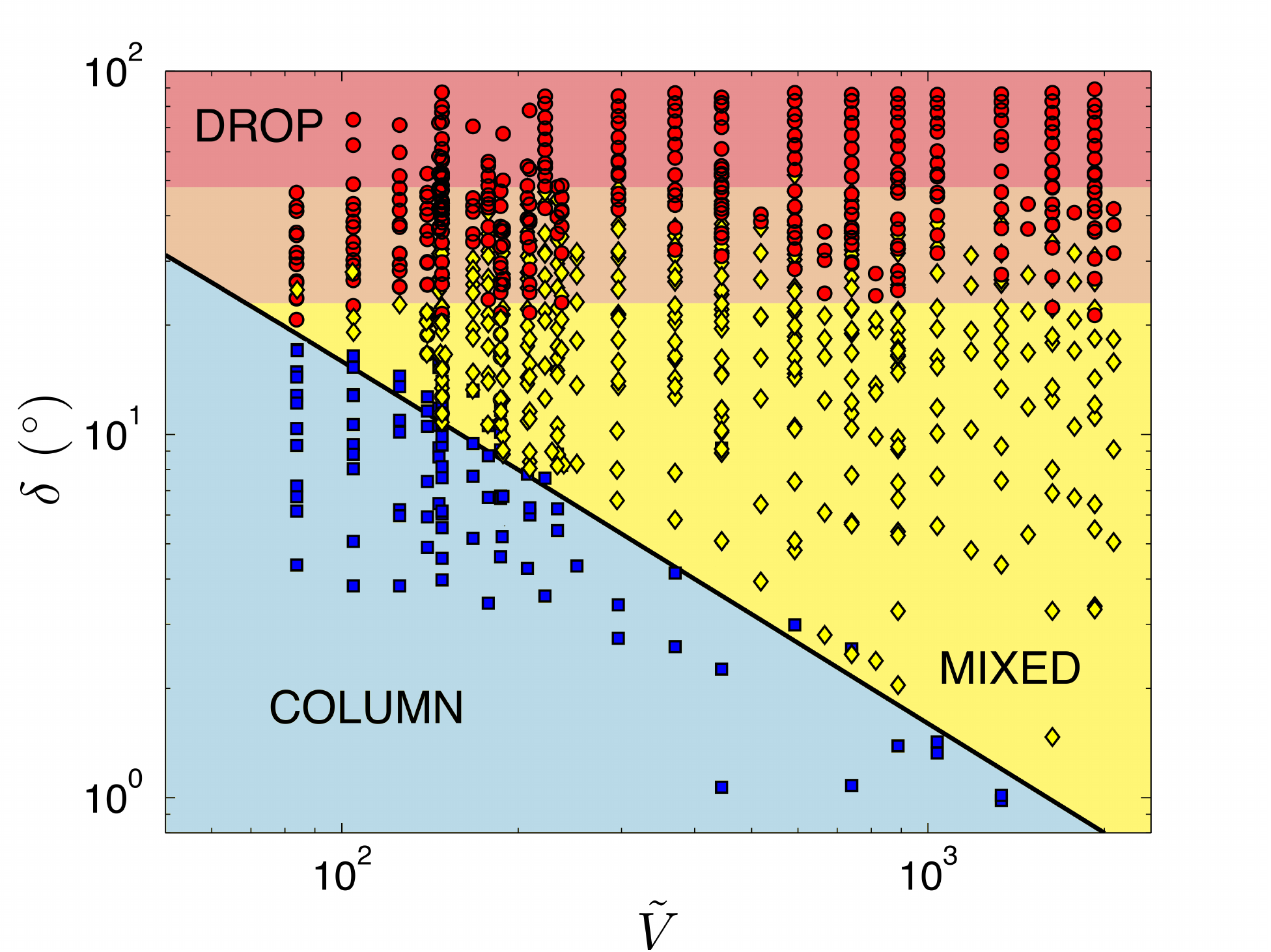}
   \caption{Morphology diagram for the touching crossed fibers ($h=0$) in the parameter space ($\tilde{V}$, $\delta$). Red circles show the drop morphology, blue squares the column morphology and yellow diamonds are the mixed morphology. The light orange region corresponds to the region where both column and mixed morphologies are observed. The solid black line corresponds to the analytical prediction of the maximum angle where the column state is possible for a given volume $\tilde{V}$.}\label{fig:Appendix_crossed}
\end{figure}

We summarize the experimental results for dodecane on touching crossed fibers in a morphology diagram of the angle between crossed fibers, $\delta$, as a function of the dimensionless volume $\tilde{V}=V/a^3$ (Fig. \ref{fig:Appendix_crossed}). These results can be compared to the analytical model for liquids with a non-zero contact angle. We observe a good agreement between the analytical prediction (black solid line) and the experimental results for the transition between the column and either mixed morphology or drop state. We observe a coexistence region where both the mixed morphology and drop state are present, which is similar to that observed for silicone oil on crossed fibers, but larger. The transitions from the mixed morphology to the coexistence region and from the coexistence region to the drop state again seem to remain independent of $\tilde{V}$.

    \bibliography{Biblio_Wetting}
    \bibliographystyle{ieeetr}

    \end{document}